# A resonance model for spontaneous cortical activity

Yanjiang Wang[1]*, Jichao Ma[1], Jiebin Luo[1], Xue Chen[2], Yue Yuan[1]

[1]*College of Control Science and Engineering, China University of Petroleum (East China), Qingdao 266580, P. R. China*

[2]*The Institute for Digital Medicine and Computer-assisted Surgery in Qingdao University, Qingdao University, Qingdao 266071, China*

*Correspondence: yjwang@upc.edu.cn

How human brain function emerges from structure has intrigued researchers for decades and numerous models have been put forward, yet none of them yields a close structure-function relation. Here we present a resonance model based on neuronal spike timing dependent plasticity (STDP) principle to describe the spontaneous cortical activity by incorporating the dynamic interactions between neuronal populations into a wave equation, which is able to accurately predict the resting brain functional connectivity (FC), including the resting-state networks. Besides, the proposed model provides strong theoretical and experimental evidences that the spontaneous dynamic coupling between brain regions fluctuates with a low frequency. Crucially, it is able to account for how the negative functional correlations emerge during resonance. We test the model with a large cohort of subjects (1038) from the Human Connectome Project (HCP) S1200 release in both time and frequency domain, which exhibits superior performance to existing eigen-decomposition models.

The human brain is highly active all the time, even being at rest, with signals propagating between sets of brain regions along structural pathways (e.g., the cortical white matters). Studies found that strong temporal correlations across different distributed brain regions during rest were observed in spontaneous fluctuations with low frequency (< 0.1Hz) measured with fMRI[1]. Yet how such slowly fluctuated and spatio-temporally organized functional patterns, referred to as resting-state functional connectivity (FC), emerge from the underlying structural connectivity (SC) measured with diffusion MRI (dMRI)[2], still remains an open question in neuroscience. Existing network-based models usually regard SC as a relatively fixed connectome or a graph network with its nodes representing brain regions and weighted edges specifying the connection strengths[3-9]. The simulated FC relies mainly on directly connected structural information such as the number of white matter tracts or connection density estimated by diffusion tractography algorithms[10], lacking the ability to describe the higher-order relationship between SC and FC. Whereas neural mass models (NMMs)[11-14] or neural field models[15,16] often suffer from the determination of a number of physiological parameters, leading to the models being less convincing. Accordingly, neither network-based models nor neural mass models are able to completely describe the overall pattern of FC.

Recent years have seen increasing studies on graph Laplacian and spectral graph analysis that bring new insights into understanding the relationship between brain structure and function[17-22], in which FC is formulated as the weighted combination of the eigenmodes of SC. These eigen-decomposition based models have been found to be able to highly capture the SC-FC relation, yet they are still unable to describe the higher-order relations between SC and FC owing to the sparseness of the graph Laplacian matrix. Crucially, none of the models is able to explain how the negative functional correlations emerge. Besides, some recent studies model the human brain functions as harmonic waves with specific spatial frequencies[23-25], in which the eigenvectors and eigenvalues of the structural connectome Laplacian respectively imply the connectome harmonics and the corresponding spatial frequencies. These harmonic-based models can merely account for the SC-FC relation qualitatively and cannot explain how the spontaneous low-frequency fluctuation cortical waves are shaped.

To remedy the above limitations, we proposed a spatiotemporal varying hypergraph Laplacian diffusion model (STV-HGLD) in a recent study[26] by incorporating the higher-order hypergraph Laplacian of SC into the regular wave equation to simulate the dynamic functional correlations, in which a sign matrix indicating the positive and negative correlations is embedded into the hypergraph Laplacian to yield the negative correlations. While the STV-HGLD model can describe the functional correlations with more accuracy, the underlying biological mechanisms remain elusive.

In fact, studies in neuroscience have evidenced that the self-organization between synaptic excitation and inhibition plays a dominant role in shaping cortical activity[27]. Particularly, resonance may occur to the structural connectome during the interplay between the local oscillations[28], nevertheless, how the resonance is shaped remains unclear. In this study, we present a resonance model to describe the spontaneous cortical activity by incorporating the dynamic interactions between neuronal populations into a wave equation. We find that the low-frequency dynamic couplings between brain regions result from the resonance with the local oscillations arising from each brain region, which fluctuate in the form of a Gaussian wave regulated by the harmonics shaped from the interplay between excitation and inhibition rather than the eigenmodes of the structural connectome *per se*. Remarkably, the resting-state FC can be theoretically determined from the model and the frequency of the fluctuation yielding the spontaneous cortical wave can be identified as well.

## Results

### Connectome dataset

The connectome dataset we used was obtained from the Human Connectome Project (HCP — www.humanconnectome.org, S1200 release)[29]. We selected subjects with high-quality fMRI data, which consists of 1038 healthy adults. The whole cerebral cortex was partitioned into 360 regions of interest (ROIs) using HCP-MMP1.0 atlas[30]. The functional connectomes were specified using Pearson's correlation between fMRI BOLD time series after data preprocessing. The structural connectomes we used were open source probabilistic connectomes[31], including 1065 subjects, in which the SC matrices of the same 1038 subjects were used. It is worth mentioning that we removed weak connections whose connection strengths are below 0.001. The mean empirical SC and FC connectomes for the first 130 subjects were demonstrated in SI Fig. 1(a) and (b). For more details, see SI Methods and ref. [31].

### Prediction of the resting-state FC

We first apply the proposed model to predict the resting-state FC. As was shown in Eq. (10) (see Methods), the dynamic couplings, $f(t)$, between brain regions at rest vary in the form of a Gaussian wave regulated by the resonance Laplacian in Eq. (15) (see Methods). Suppose the sign matrix $S$ is known, $C$ can be initially set to be 1, and $\beta$, the decay factor, which controls the time width of $f(t)$, is experimentally chosen as 0.1, then the theoretical dynamic functional connectivity can be obtained.

The performances were assessed for each subject respectively with the Pearson correlation between $f(t)$ and the empirical FC. Figure 1 demonstrates the overall performance of the proposed model in predicting FC. Figure 1 (a) shows the evolution of the Pearson correlations for the first 130 subjects over time, respectively. Figure 1 (b) plots the maximum Pearson correlation values (characterized by $f(t)$ at the critical time) for all the 1038 subjects. Besides, the performances of the models based on graph diffusion (GD)[17, 18] and hypergraph diffusion (HGD)[32] are also graphically displayed. It can be obviously observed that very high Pearson correlation values (> 0.8) are obtained for almost all the subjects using the proposed model, which significantly exceeds the performance of HGD and GD models. For clear comparison, Fig. 1 (c) shows the mean predicted FC connectome over the first 130 subjects and Fig. 1 (d) demonstrates the predicted FC using seed-based surface plots, both of which bear strong resemblance to the empirical FC measured with fMRI.

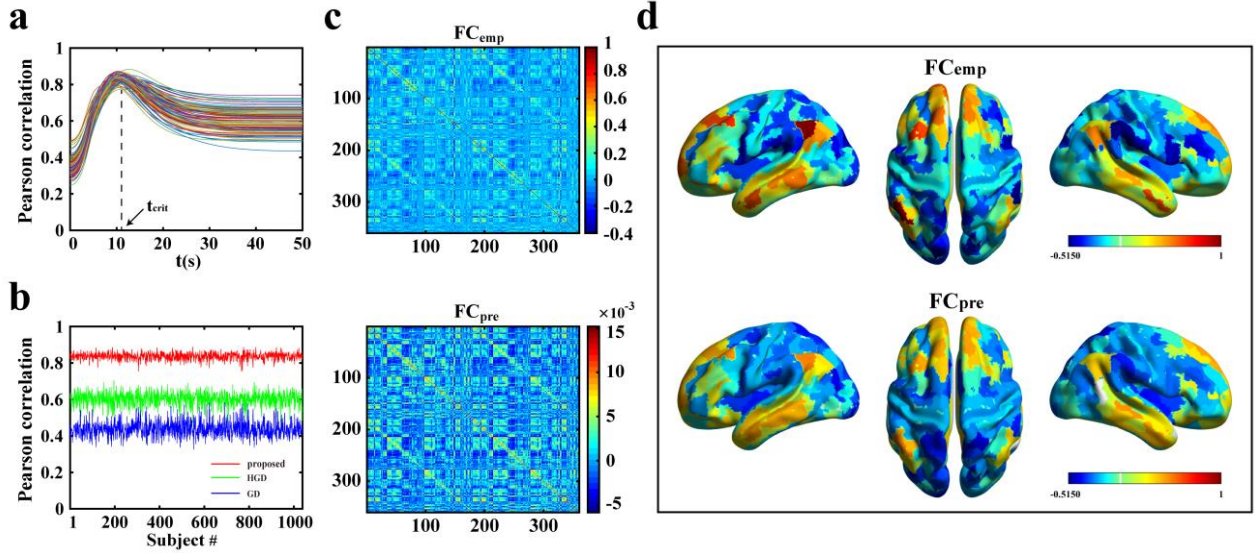

**Fig. 1** The overall performance of the proposed model in predicting the resting-state FC. (a) The evolution of the Pearson correlations with the empirical FC for the first 130 subjects over time ($p<1.0e-6$). The maximum correlations arise at 10.8115±0.9092s; (b) The maximum Pearson correlation values for all the 1038 subjects using the proposed, HGD, and GD models, with the mean correlations being 0.8366±0.0198, 0.6026±0.0370, and 0.4383±0.0426, respectively ($p<1.0e-6$); (c) The comparison between the mean predicted FC and the mean empirical FC over the first 130 subjects. The Pearson correlation between the two is 0.9028 ($p<1.0e-6$); (d) The comparison between the surface plots of seed-based FC (the first subject, both hemispheres). The color maps are produced using BrainNet Viewer[33] with seeding at left area PFm Complex (area 149 in the 360-ROI atlas) in the brain network. Cold/warm color indicates negative/positive functional connections

**Prediction of 20 resting-state networks (RSNs)**

To show the predictive power of our model, we further perform evaluations on predicting the resting-state networks (RSNs). For each hemisphere of the 180 cortical regions, we extract 10 RSNs[31], namely, visual, somatomotor, cingulo-opercular, dorsal attention, language, frontoparietal, auditory, default model, multimodal, and orbito-affective, respectively. The Pearson correlations for each of the predicted RSNs are listed in Table I. The comparison between the mean predicted RSNs and the mean empirical RSNs over the first 130 subjects are demonstrated in Fig. 2. It can be clearly seen that higher Pearson correlations are also obtained for all the 20 RSNs, validating that our model is able to predict not only the global pattern of FC but also local RSNs.

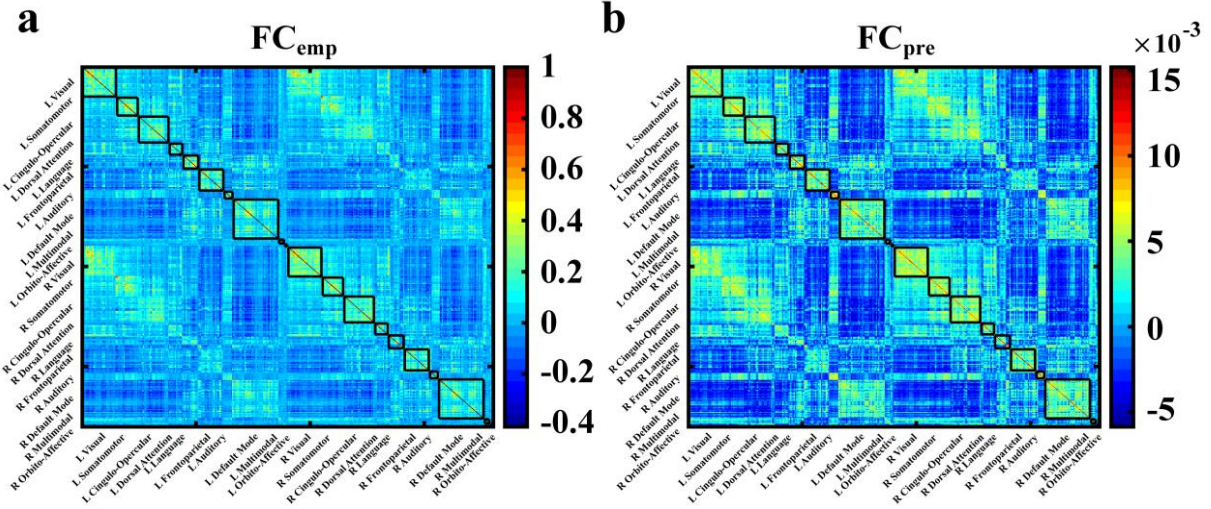

**Fig. 2** The comparison between the mean predicted RSNs and the mean empirical RSNs for the first 130 subjects. (**a**) The mean empirical RSNs. (**b**) The corresponding mean predicted RSNs. For clearly visualization, the brain areas are reordered and organized into 10 RSNs for each hemisphere. See ref. [31] for more details.

**TABLE I.** The Pearson correlations for each of the 10 predicted RSNs ($p<1.0e-6$) (LH — Left Hemisphere; RH — Right Hemisphere)

| RSNs | LH | RH |
|---|---|---|
| Visual | 0.7913±0.0545 | 0.7868±0.0595 |
| Somatomotor | 0.7563±0.0679 | 0.7496±0.0685 |
| Cingulo-opercular | 0.7374±0.0428 | 0.7544±0.0509 |
| Dorsal attention | 0.8350±0.0449 | 0.8227±0.0370 |
| Language | 0.8508±0.0380 | 0.8672±0.0313 |
| Frontoparietal | 0.8436±0.0394 | 0.8141±0.0337 |
| Auditory | 0.7671±0.0778 | 0.8030±0.0463 |
| Default mode | 0.7398±0.0420 | 0.7609±0.0367 |
| Multimodal | 0.8617±0.0437 | 0.8553±0.0503 |
| Orbito-Affective | 0.9207±0.0467 | 0.9158±0.0448 |

**Obtaining FC from SC directly**

It is worth highlighting that, similar to the previous methods, the aforementioned predicted FC is evaluated by searching for the maximum Pearson correlations (Fig. 1(a)), which draws on the empirical FC to determine the critical time, $t_{crit}$. Surprisingly, we find that the critical time can also be estimated by computing the Pearson correlation between $f(t)$ and SC. As shown in Fig. 3 (a), although the Pearson correlations with SC are much lower than with FC, the peak values can also be observed on the correlation curves. Thus, we can obtain the FC directly after the critical time is determined, suggesting that our model is able to theoretically yield the resting-state brain FC from SC given that the connection strength and signs are known. It is worth mentioning that the Pearson correlations for the directly computed FC show slight drop because, for each of the subjects, there is a little variance between the optimal critical time with FC (Fig. 1 (a)) and the estimated critical time with SC (Fig. 3 (a)).

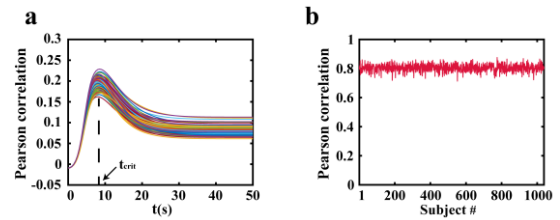

**Fig. 3** Obtaining FC from SC directly. (a) The evolution of the Pearson correlations with empirical SC for the first 130 subjects over time ($p<1.0e-6$); (b) The maximum Pearson correlation values (characterized by $f(t)$ at the critical time in (**a**)) for all the 1038 subjects. The average correlation is 0.8038±0.0247 ($p<1.0e-6$).

**Frequency analysis of the spontaneous cortical wave**

In what follows, we evaluate the performance of the proposed model in frequency domain to show how the frequency varies during the propagation of the spontaneous cortical wave. Figure 4 (a) demonstrates the evolution of the Pearson correlations between $F(f_r)$ ( see Eq. (17) in Methods), and the empirical FC for the first 130 subjects over frequency, respectively. It can be clearly observed that for every subject, when resonance occurs, the correlation reaches the maximum at frequency around $f_r = 0.0529$Hz, fitting well to the BOLD fluctuations (<0.1Hz).

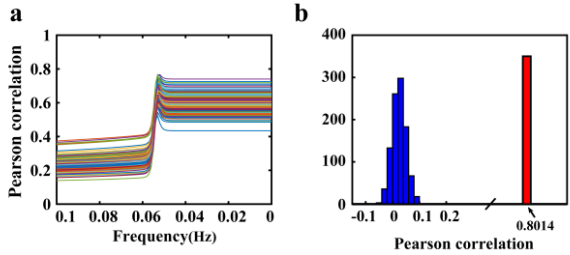

**Fig. 4** Frequency and robustness analysis of the model. (a) The evolution of the Pearson correlations with empirical FC for the first 130 subjects over frequency. The maximum correlations arise at frequency around 0.0529Hz ($p<1.0e-6$); (b) The histogram of the Pearson correlations between the mean empirical FC and the mean predicted FC (blue) as well as the mean predicted FC without permuting (red) over the first 130 subjects;

In particular, when $f_r = 0$, we can compute the steady state theoretically, i.e., $F(0) = \sqrt{\pi/\beta L_R}$ or $\int_{-\infty}^{+\infty} e^{-\beta L_R t^2} dt = \sqrt{\pi/\beta L_R}$, indicating that $f(t)$ will keep constant after resting-state FC arises, from which the Pearson correlations tend to be stable when the fluctuation tapers off (see Fig. 1 (a) and Fig. 4 (a)).

**Robustness analysis**

To show the robustness of our model in predicting FC, we first randomly permute the mean SC with 1000 times, and then we compute the FC using the model for each randomly generated SC matrix. Figure 4 (b) shows the histogram of the Pearson correlations between the mean empirical FC and the mean predicted FC as well as the mean predicted FC without permuting for the first 130 subjects, highlighting that only the predicted FC using the observed SC is significant while those obtained from the randomized SC matrices are negligible.

**Discussion**

In this work, we demonstrate a new model to describe the spontaneous cortical activity by incorporating the dynamic interactions between neuronal regions into a wave equation. Theoretical solution shows that the dynamic couplings between brain regions fluctuate in the form of a Gaussian wave regulated by the harmonics shaped from the interplay between excitation and inhibition, which may lead to resonance with the structural connectome at a critical time from which the resting-state FC arises. The main contributions of our work are as follows.

Firstly, different from previously reported findings based on eigen-decomposition[17,18,20-25], in which the eigenvectors of the structural connectome Laplacian are modeled as the fundamental building blocks of the functional network, our modeling results reveal that the resting brain function is more likely dominated by two essential factors pertaining to SC, one is the Gaussian decay of the graph Laplacian eigenvalues of the structural connectome, which controls the fluctuation of the spontaneous cortical wave with low frequency; the other is the sign of the connections (excitatory or inhibitory) between brain regions, which shapes the harmonics with each frequency components. Our model can directly and accurately predict the resting-state FC given the above two factors, which produces close functional patterns with the empirical FC, without tuning any physiological parameters (Fig.1 (c), (d)).

Secondly, our model provides strong theoretical and experimental evidences that the spontaneous dynamic coupling between brain regions fluctuates with a low frequency. We analyze our model in frequency domain and find that the predicted resting-state FC really emerges at a low frequency (around 0.0529Hz), which is highly consistent with the BOLD signals measured with fMRI. This finding can be confirmed by exploring the relationship between the number of sign changes (zero crossings) of each harmonic and the SC Laplacian eigenvalues without or with Gaussian decay at the corresponding critical time when FC emerges. As shown in Fig. 5, most harmonics with higher frequency correspond to higher SC Laplacian eigenvalues (Fig.5 (a)) but to lower eigenvalues after Gaussian decay, thereby causing the overall spontaneous cortical wave fluctuate with a lower frequency.

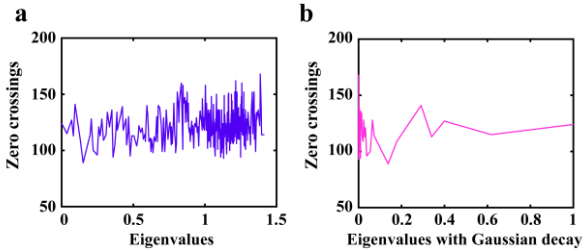

**Fig. 5** The relationship between the number of zero crossings (sign changes) of each harmonic and the SC Laplacian eigenvalues without or with the Gaussian decay at the corresponding critical time when FC emerges. (**a**) Zero crossing changes with the SC Laplacian eigenvalues; (**b**) Zero crossing changes with the eigenvalues after Gaussian decay at the corresponding critical time when FC emerges.

Thirdly, our model is able to account for how the negative functional correlations between brain regions emerge during rest. To the best of our knowledge, we are the first to apply the neuronal spike timing dependent plasticity (STDP) principle[33,34] to unravel how anti-correlations arise during the interplay of excitation and inhibition between brain regions. Although, currently, we are unable to obtain the relevant information due to the inability of the current diffusion imaging and tractography techniques to measure the direction and causality between connections, we have proved that the predicted FC shows nearly the same signs as empirical FC by embedding the sign matrix extracted from the measured FC into the wave equation directly (see Eq. (13)).

Fourthly, our model outperforms some state-of-the-art eigen-decomposing based models in predicting resting-state FC. One of the typical models is the graph diffusion (GD) model put forward by Abdelnour et al.[17, 18], which is able to describe the dynamic functional correlations between brain regions using the graph Laplacian of SC. The model is linear and takes the form of an exponential decay, which is only capable of modeling a portion of FC due to the sparseness of the SC Laplacian. To address this issue, we extended the graph diffusion model to a hypergraph diffusion (HGD) model in a recent study[32], albeit raising the accuracy, the average Pearson correlation is still much lower than the proposed model (Fig. 1 (b)).

Another typical model using spectral graph mapping proposed by Becker et al.[19], in which a connectome dataset with 44 subjects and 360 ROIs using the same HCP data was used to evaluate their model. The highest Pearson correlation value with 0.5178 was obtained for group spectral mapping with half of the subjects for training. Recently, Tewarie et al.[22] explored the SC-FC mapping using series expansion and eigenmode respectively, in which a connectome dataset with 10 subjects and 78 ROIs using the same HCP data was used to evaluate their approach. The highest Pearson correlation value of 0.62 was achieved with eigenmode and 0.6 with series expansion, respectively. Besides, a recent model based on connectome embedding algorithm and deep learning was proposed by Rosenthal et al.[36] for modeling the higher-order SC-FC relations, in which a connectome dataset with 100 subjects and 82 ROIs using the same HCP data was used to test their model. The maximum Pearson correlation value of 0.52 is obtained for direct connections and 0.6 for indirect connections in the testing set.

Our proposed model can achieve higher Pearson correlation as opposed to the above models, as aforementioned, the average Pearson correlation is over 0.8366 across all the 1038 subjects for both direct connections and indirect connections. In addition, our model requires no training as the above models.

Taken together, the proposed resonance model is able to bring to light how neural activities are coupled and propagate in brain. Importantly, it may give us hope to conjecture whether the same way is carried out in brain during performing cognitive tasks such as learning, memory, as well as decision making[37].

## Methods

### Dynamic brain network notation

The dynamic brain network in this work is defined as $G=(V,E,S)$, where $V=\{v_i|i\in 1,2,\cdots,n\}$ represents $n$ brain regions, $E=\{(i,j)|i\in V, j\in V\}$ labels the edge linking region $i$ and $j$, whose weight $w_{i,j}$ measuring the anatomical connection strength between the two regions. $W=\{w_{i,j}|(i,j)\in E\}$ denotes the structural connectivity matrix, and $S=\{s_{i,j}|(i,j)\in E\}$ defines the interplay of excitation and inhibition between region $i$ and region $j$. For simplicity, here we just denote by $s_{i,j}$ the signs of the connections between the two brain regions after a period of time, with $s_{i,j}=1$ and $s_{i,j}=-1$ respectively specifying the excitatory and inhibitory connections. Let $D$ denote the degree matrix, $D=diag(d_i)$, $d_i=\sum_{j=1}^{n}w_{i,j}$, then the graph Laplacian of $G$ is defined as $L=D-W$ and $L$ can be decomposed as $L=U\Lambda U^T$, where $U$ and $\Lambda$ are the matrices containing the eigenvectors and eigenvalues of $L$, respectively.

### Spontaneous cortical activity modeling

Studies in neuroscience show that if a firing neuron causes another neuro to fire, the connection strength between the two neurons will increase otherwise decrease, which is referred to as spike timing dependent plasticity (STDP)[34]. The STDP can be generalized to the macroscale case, i.e., the connection strength between two brain regions will also fluctuate during information exchange between them, suggesting that if the connection between two regions exhibits excitatory behavior, the connection strength will increase, otherwise decrease for an inhibitory connection.

Specifically, supposing $x=[x_1, x_2, \cdots, x_n]^T$, $x_i = x(i,t)$ specifies the cortical activity signal of the $i$th brain region at time $t$, $T$ indicates transpose.

Regarding the principle of STDP and the above brain network notation, the firing of the neurons in the $j$th brain region will cause excitatory or inhibitory effect to the $i$th brain region, giving rise to a little fluctuation to the weight $w_{i,j}$ vary with time, i.e., $\Delta w_{i,j}(t)$, which usually varies linearly within a short period of time[35], i.e.,

$$\Delta w_{i,j}(t)=w_{i,j}t \quad (1)$$

Then the total first-order fluctuation on the $i$th brain region caused by all the other brain regions linking to it can be expressed as

$$\nabla x_i = \sum_{j\in n}\Delta w_{i,j}(t)\left(s_{i,j}x_j - x_i\right) \quad (2)$$

where $s_{i,j}$ indicates the sign of the connection (excitatory or inhibitory) between region $i$ and $j$. Bring all the regions together, the above first-order dynamics can be described as

$$\nabla x_1 = \left(w_{1,1}s_{1,1}x_1 + w_{1,2}s_{1,2}x_2 + \cdots + w_{1,n}s_{1,n}x_n\right)t - \left(w_{1,1}x_1 + w_{1,2}x_1 + \cdots + w_{1,n}x_1\right)t$$
$$\nabla x_2 = \left(w_{2,1}s_{2,1}x_1 + w_{2,2}s_{2,2}x_2 + \cdots + w_{2,n}s_{2,n}x_n\right)t - \left(w_{2,1}x_2 + w_{2,2}x_2 + \cdots + w_{2,n}x_2\right)t$$
$$\vdots$$
$$\nabla x_n = \left(w_{n,1}s_{n,1}x_1 + w_{n,2}s_{n,2}x_2 + \cdots + w_{n,n}s_{n,n}x_n\right)t - \left(w_{n,1}x_n + w_{n,2}x_n + \cdots + w_{n,n}x_n\right)t$$
$$(3)$$

Using matrix notations, Eq. (3) can be rewritten as

$$\nabla x = \begin{bmatrix}\nabla x_1 \\ \nabla x_2 \\ \vdots \\ \nabla x_n\end{bmatrix} = \begin{bmatrix} w_{1,1}s_{1,1} & w_{1,2}s_{1,2} & \cdots & w_{1,n}s_{1,n} \\ w_{2,1}s_{2,1} & w_{2,2}s_{2,2} & \cdots & w_{2,n}s_{2,n} \\ & & \ddots & \\ w_{n,1}s_{n,1} & w_{n,2}s_{n,2} & \cdots & w_{n,n}s_{n,n}\end{bmatrix}\begin{bmatrix}x_1 \\ x_2 \\ \vdots \\ x_n\end{bmatrix}t$$

$$-\begin{bmatrix}(w_{1,1}+w_{1,2}+\cdots+w_{1,n}) & 0 & \cdots & 0 \\ 0 & (w_{2,1}+w_{2,2}+\cdots+w_{2,n}) & \cdots & 0 \\ & & \ddots & \\ 0 & 0 & \cdots & (w_{n,1}+w_{n,2}+\cdots+w_{n,n})\end{bmatrix}\begin{bmatrix}x_1 \\ x_2 \\ \vdots \\ x_n\end{bmatrix}t$$

$$= \begin{bmatrix} w_{1,1} & w_{1,2} & \cdots & w_{1,n} \\ w_{2,1} & w_{2,2} & \cdots & w_{2,n} \\ & & \ddots & \\ w_{n,1} & w_{n,2} & \cdots & w_{n,n}\end{bmatrix} \circ \begin{bmatrix} s_{1,1} & s_{1,2} & \cdots & s_{1,n} \\ s_{2,1} & s_{2,2} & \cdots & s_{2,n} \\ & & \ddots & \\ s_{n,1} & s_{n,2} & \cdots & s_{n,n}\end{bmatrix}\begin{bmatrix}x_1 \\ x_2 \\ \vdots \\ x_n\end{bmatrix}t$$

$$-\begin{bmatrix} d_1 & 0 & \cdots & 0 \\ 0 & d_2 & \cdots & 0 \\ & & \ddots & \\ 0 & 0 & \cdots & d_n \end{bmatrix} \begin{bmatrix} x_1 \\ x_2 \\ \vdots \\ x_n \end{bmatrix} t$$

$$= (\boldsymbol{W} \circ \boldsymbol{S} - \boldsymbol{D}) t \boldsymbol{x} = -(\boldsymbol{D} - \boldsymbol{W} \circ \boldsymbol{S}) t \boldsymbol{x} \quad (4)$$

where $\boldsymbol{W}$, $\boldsymbol{D}$, and $\boldsymbol{S}$ denote the adjacency, degree, and the sign matrices respectively, and '$\circ$' refers to the Hadamard product. Using the definition of graph Laplacian, here we define $\boldsymbol{L}_R = \boldsymbol{D} - \boldsymbol{W} \circ \boldsymbol{S}$, namely the resonance Laplacian.

Applying the $\nabla$ operator again, we obtain the second-order fluctuation, i.e., the Laplace operator, as follows:

$$\nabla^2 \boldsymbol{x} = \nabla(\nabla \boldsymbol{x}) = (-\boldsymbol{L}_R t)(-\boldsymbol{L}_R t \boldsymbol{x}) = \boldsymbol{L}_R^2 t^2 \boldsymbol{x} \quad (5)$$

Substituting Eq. (5) into the following regular wave equation:

$$\nabla^2 \boldsymbol{x} = \frac{1}{\beta^2} \frac{\partial^2 \boldsymbol{x}}{\partial t^2} \quad (6)$$

where $\beta$ is the decay factor.

Then we obtain

$$\boldsymbol{L}_R^2 t^2 \boldsymbol{x} = \frac{1}{\beta^2} \frac{\partial^2 \boldsymbol{x}}{\partial t^2} \quad (7)$$

Equation (7) has an explicit solution (see SI Note 1 for the proof)

$$\boldsymbol{x} = k e^{-\frac{1}{2}\beta \boldsymbol{L}_R t^2} \quad (8)$$

Applying the eigen-decomposition of $\boldsymbol{L}_R$, we have

$$\boldsymbol{x} = k \cdot \boldsymbol{U}_R \cdot e^{-\frac{1}{2}\beta \Lambda_R t^2} \cdot \boldsymbol{U}_R^T \quad (9)$$

Then the dynamic functional correlation is given as

$$f(t) = \boldsymbol{x} \cdot \boldsymbol{x}^T = C e^{-\beta \boldsymbol{L}_R t^2} \quad (10)$$

where $C = k^2$ is a constant.

**The determination of $\boldsymbol{L}_R$ at resonance**

Since the sign matrix $\boldsymbol{S}$ is equivalent to $\boldsymbol{W}_H \circ \boldsymbol{S}$, where $\boldsymbol{W}_H$ corresponds to an adjacency matrix with all the elements being 1 excluding the diagonal entries, then $\boldsymbol{L}_R$ can be rewritten as

$$\boldsymbol{L}_R = \boldsymbol{D} - \boldsymbol{W} \circ \boldsymbol{W}_H \circ \boldsymbol{S} \quad (11)$$

Factorizing matrix $\boldsymbol{L}_R$ using eigen-decomposition, i.e., $\boldsymbol{L}_R = \boldsymbol{U}_R \Lambda_R \boldsymbol{U}_R^T$, where $\boldsymbol{U}_R$ represents the harmonics at resonance and $\Lambda_R$ regulates the frequencies at resonance.

Suppose the self-organizing interaction of excitation and inhibition between brain regions will lead to resonance with SC at a critical time $t_{crit}$ where the correlation between $f(t)$ and the observed FC reaches a maximum.

Factorizing the FC connectome matrix using eigen-decomposition, we obtain,

$$\text{FC} = \boldsymbol{U}_F \Lambda_F \boldsymbol{U}_F^{-1} \quad (12)$$

Without loss of generality, we set $k$ to 1. It can be inferred from Eqs. (9), (10), and (12), to make the correlation between $f(t)$ and FC maximized at $t_{crit}$, the following two relations should hold, i.e., $\boldsymbol{U}_F \approx \boldsymbol{U}_R$, $\Lambda_F \approx e^{-\beta \Lambda_R t_{crit}^2}$. Note that from $\boldsymbol{L}_R = \boldsymbol{D} - \boldsymbol{W} \circ \boldsymbol{S}$, it can be observed that the signs of the elements in $-\boldsymbol{L}_R$ are the same as those in the sign matrix $\boldsymbol{S}$ except the diagonal entries. Applying Tailor expansion gives

$$f(t) = e^{-\beta \boldsymbol{L}_R t^2} \approx \boldsymbol{I} - \beta \boldsymbol{L}_R t^2 \quad (13)$$

where $\boldsymbol{I}$ denotes the identity matrix. This approximation implies that the signs of the off-diagonal entries in $f(t)$ are nearly the same as those in $\boldsymbol{S}$, suggesting that we can make the sign of each elements in matrix $\boldsymbol{S}$ equal to those of the corresponding elements in FC. Thus $\boldsymbol{U}_R$ can be modeled as the harmonics yielded by the following Laplacian eigen-decomposition

$$\boldsymbol{L}_H = (\boldsymbol{D}_H - \boldsymbol{W}_H) \circ \boldsymbol{S} = \boldsymbol{U}_H \Lambda_H \boldsymbol{U}_H^T \quad (14)$$

where $\boldsymbol{W}_H$ corresponds to an adjacency matrix with all the elements being 1 excluding the diagonal entries, $\boldsymbol{D}_H$ is the degree matrix of $\boldsymbol{W}_H$.

As for $\Lambda_R$, our previous experimental findings have already shown that the relation between the eigenvalues of FC, $\Lambda_F$, and the Laplacian eigenvalues of SC, $\Lambda$, conforms to a negative exponential relation [32], i.e., $\Lambda_F \approx e^{-\alpha \Lambda}$, whereas $\Lambda_F$ and $\Lambda_H$ are not exponential correlated, therefore $\Lambda_R$ can be replaced with $\Lambda$, where $\alpha$ is a constant.

Taken together, the resonance Laplacian, $L_R$, can be reformulated as the following form when resonance occurs,

$$L_R = U_H \Lambda U_H^T \quad (15)$$

where $U_H$ represents the harmonics shaped by the self-organizing interplay of excitation and inhibition between brain regions (determined by $W_H$ and $S$), while $\Lambda$ specifies the natural frequencies of SC (depend on $W$).

### Analyzing $f(t)$ in frequency domain

Since $\int_{-\infty}^{+\infty} e^{-ax^2} dx = \sqrt{\pi/a}$ $(a > 0)$, let $a = \beta L_R$, we can obtain the Fourier transform of $f(t)$, $F(\omega)$, as follows (see SI Note 2 for the proof),

$$F(\omega) = \int_{-\infty}^{+\infty} f(t) e^{-j\omega t} dt = \sqrt{\frac{\pi}{\beta L_R}} e^{-\frac{\omega^2}{4\beta L_R}} \quad (16)$$

Given $\omega = 2\pi f_r$, where $f_r$ stands for the real frequency (Hz), then we have

$$F(f_r) = \sqrt{\frac{\pi}{\beta L_R}} e^{-\frac{\pi^2 f_r^2}{\beta L_R}} \quad (17)$$

### Data availability

The connectome dataset we used was obtained from the Human Connectome Project (HCP — www.humanconnectome.org, S1200 release). The probabilistic structural connectome data was shared by B. Q. Rosen and E. Halgren at http://zenodo.org/record/4060485. The rs-fMRI data were provided by the Human Connectome Project, WU-Minn Consortium (Principal Investigators: David Van Essen and Kamil Ugurbil; 1U54MH091657) funded by the 16 NIH Institutes and Centers that support the NIH Blueprint for Neuroscience Research; and by the McDonnell Center for Systems Neuroscience at Washington University.

### Code availability

The code was written in Matlab and can be downloaded at

# Acknowledgements

Yanjiang Wang appreciates Richard M. Shiffrin and Olaf Sporns for their invaluable help and discussions on the research during his visit to Indiana University, Bloomington. The authors also cordially thank Farras Abdelnour for sharing the implementation codes of their network diffusion model. This work was funded in part by the National Natural Science Foundation of P.R. China (Grant No. 62072468).

## Author contributions

Y. W. and J. M. contributed equally. Y. W. conceived the study, conducted the experiments and wrote the manuscript; J. Ma. designed the algorithms and performed the analysis; J. L., X. C, and Y. Y. prepared and processed the MRI data, conducted the experiments, and drew the figures; and all the authors checked the manuscript.

## Competing interests

The authors declare no competing interests.

## Additional information

**Supplementary information** is available for this paper at

**Competing financial interests**: The authors declare no competing financial interests.

**Reprints and permission information** is available at

# Supplementary Information for "A Resonance Model for Spontaneous Cortical Activity"

## Supplementary Methods

**Data Acquisition**

The MRI data we used was obtained from the Human Connectome Project (HCP —www.humanconnectome.org, S1200 release). We selected subjects with high-quality fMRI data, which consists of 1038 adults.

For the structural and diffusion data: All Human Connectome Project imaging data[1] were acquired on a Siemens Skyra 3T scanner with a customized SC72 gradient insert. T1w 3D MPRAGE were acquired with TR=2400ms, TE=2.14ms, TI=1000ms, flip angle=8°, FOV=224x224, 0.7mm isotropic voxel, bandwidth=210Hz/px, iPAT=2, Acquisition time=7:40 (min:sec). Diffusion weighted images were acquired with Spin-echo EPI sequences (b-values=0, 1000, 2000, 3000 s/mm2 in approximately 90 gradient directions, TR=5520 ms, TE=89.5 ms, flip angle=78°, refocusing flip angle=160°, FOV=210x180 (RO x PE) matrix=168x144 (RO x PE), slice thickness=1.25mm, 111 slices, 1.25mm isotropic voxels, Multiband factor=3, Echo spacing=0.78ms, BW=1488 Hz/Px, Phase partial Fourier 6/8[2].

For the resting-state fMRI: Sequence: Gradient-echo PLI, TR: 720ms, TE: 33.1ms, flip angle 52 deg, FOV: 208x180mm (RO x PE), Matrix: 104x90 (RO x PE), Slice thickness: 2.0 mm; 72 slices; 2.0 mm isotropic voxels. Multiband factor: 8, Echo spacing: 0.58 ms, BW: 2290 Hz/Px. Resting state data produced 1200 frames per run (time points), of total duration: 14 minutes 33 seconds. For further details, see ref. [3].

**Extraction of Structural and Functional Connectomes**

**Probabilistic Structural Connectome**

We used open source probabilistic connectomes[9]. The connectomes were constructed by probabilistic tractography and the nodes were also defined by HCP-MMP1.0 atlas. The probabilistic connectomes have been normalized and symmetric.

**fMRI Data Preprocessing and Generating Functional Connectome**

All resting-state fMRI data (two runs) were minimally preprocessed with echo planar imaging gradient distortion correction, motion correction, field bias correction, spatial transformation and normalization into a common MNI space[4] and artifact removal using independent component analysis (ICA) + FIX[5,6]. For each fMRI imaging run, we first used dpabi (http://www.rfmri.org/dpabi)[7] to remove global signal and estimated the functional connectomes using Pearson's correlation between time points for each of the 1038 subjects, then we average the two different functional connectomes of each subject. The network nodes were defined using HCP-MMP1.0 atlas[8].

The mean empirical SC and FC connectomes over the first 130 subjects are demonstrated in SI Fig. 1(a) and (b), respectively.

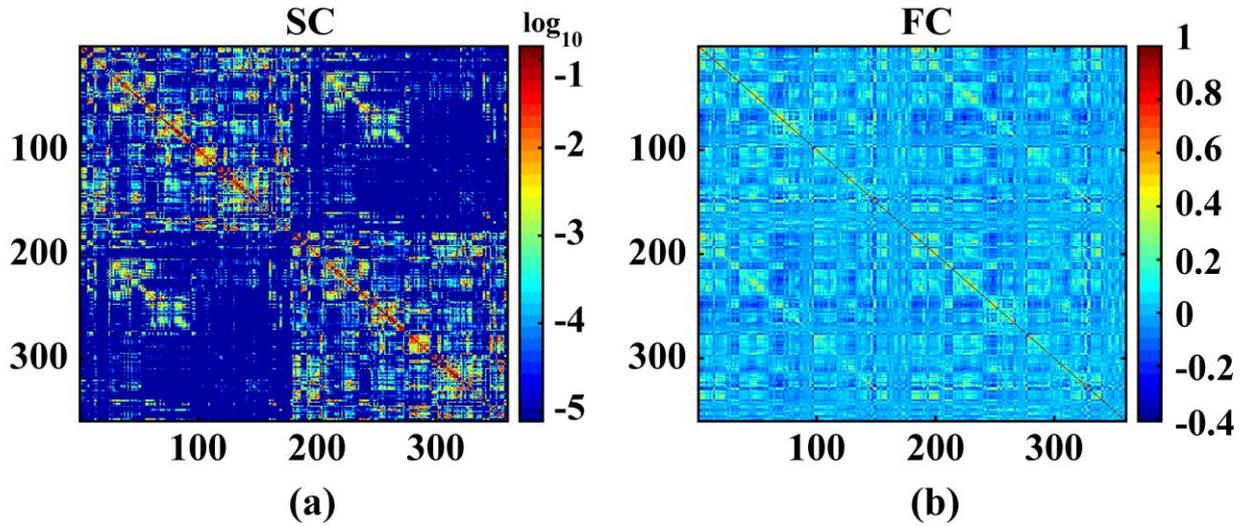

**Supplementary Fig. 1.** The mean empirical SC and FC connectomes derived from the HCP connectome dataset over the first 130 subjects. **(a)** The mean empirical SC connectome; **(b)** The corresponding mean empirical FC connectome. Both matrices are symmetric and the matrices' elements are arranged such that the upper left quadrant maps the left hemisphere, the lower right quadrant the right hemisphere, and the off-diagonal quadrants the interhemispheric connections. For clear visualization, the connections in SC are displayed with a logarithmic color map.

## Supplementary Note 1: Proof of The Solution of The Wave Equation

As described in the main text, the wave equation can be expressed as

$$\frac{\partial^2 \boldsymbol{x}}{\partial t^2} = \beta^2 \boldsymbol{L}_R^{\ 2} t^2 \boldsymbol{x} \tag{1}$$

The solution is derived as follows.

Assume $\frac{\partial \boldsymbol{x}}{\partial t} = \boldsymbol{\rho}$, then $\frac{\partial^2 \boldsymbol{x}}{\partial t^2} = \frac{\partial \boldsymbol{\rho}}{\partial t} = \frac{\partial \boldsymbol{\rho}}{\partial \boldsymbol{x}} \cdot \frac{\partial \boldsymbol{x}}{\partial t} = \boldsymbol{\rho} \cdot \frac{\partial \boldsymbol{\rho}}{\partial \boldsymbol{x}}$, and then the wave equation can be rewritten as

$$\rho \frac{\partial \rho}{\partial x} = \beta^2 L_R^2 t^2 x \qquad (2)$$

i.e., $\rho \partial \rho = \beta^2 L_s^2 t^2 x \partial x$, integrating on both sides gives,

$$\frac{1}{2}\rho^2 = \frac{1}{2}\beta^2 L_R^2 t^2 x^2$$

Considering the damping effect during the propagating of neural activity signals, the solution $\rho$ can be expressed as

$$\rho = -\beta L_R t x \qquad (3)$$

i.e., $\frac{\partial x}{\partial t} = -\beta L_R t x$, shifting $x$ to the left side and $t$ to the right side, we have

$$\frac{\partial x}{x} = -\beta L_R t \partial t \qquad (4)$$

Integrating on both sides gives

$$ln(x) = -\frac{1}{2}\beta L_R t^2 + k \qquad (5)$$

Then we obtain

$$x = k e^{-\frac{1}{2}\beta L_R t^2} \qquad (6)$$

where $k$ is a constant.

## Supplementary Note 2: Proof of The Fourier Transform of $f(t)$

Since $\int_{-\infty}^{+\infty} e^{-at^2} dx = \sqrt{\frac{\pi}{a}}$ $(a > 0)$, then

$$F(\omega) = \int_{-\infty}^{+\infty} f(t) e^{-j\omega t} dt = \int_{-\infty}^{+\infty} e^{-at^2} e^{-j\omega t} dt$$

$$= \int_{-\infty}^{+\infty} e^{-(at^2 + j\omega t)} dx = \int_{-\infty}^{+\infty} e^{-a\left(t + \frac{j\omega}{2a}\right)^2 - \frac{\omega^2}{4a}} dt$$

$$= e^{-\frac{\omega^2}{4a}} \int_{-\infty}^{+\infty} e^{-\left(\sqrt{a}t + \frac{j\omega}{2\sqrt{a}}\right)^2} dt \qquad (7)$$

Let $u = \sqrt{a}t + \frac{j\omega}{2\sqrt{a}}$, then

$$F(\omega) = e^{-\frac{\omega^2}{4a}} \int_{-\infty}^{+\infty} e^{-u^2} \frac{1}{\sqrt{a}} du$$

$$= \frac{1}{\sqrt{a}} e^{-\frac{\omega^2}{4a}} \int_{-\infty}^{+\infty} e^{-u^2} du$$

$$= \sqrt{\frac{\pi}{a}} e^{-\frac{\omega^2}{4a}} \tag{8}$$

Let $a = \beta L_R$, we arrive at

$$F(\omega) = \sqrt{\frac{\pi}{\beta L_R}} e^{-\frac{\omega^2}{4\beta L_R}} \tag{9}$$

## Supplementary References